\documentclass[12pt]{article}
\makeatletter

\def\@authoraddress{}
\def\@title{}
\def\title#1{\gdef\@title{{\par\vskip-10pt\Large\bf
\baselineskip20pt\centering\ignorespaces\uppercase{#1}\vskip6pt}}%
\setcounter{table}{0}      \setcounter{figure}{0}
\setcounter{equation}{0}   \setcounter{section}{0}
\setcounter{subsection}{0} \setcounter{subsubsection}{0}
\setcounter{paragraph}{0}
}

\def\authors#1{\expandafter\def\expandafter\@authoraddress\expandafter
{\@authoraddress %
{\dimen0=-\prevdepth \advance\dimen0 by1.5\baselineskip
\nointerlineskip \centering
\vrule height\dimen0 width0pt\relax\ignorespaces\large\sc#1\par
}%
}%
}

\def\addresses#1{\expandafter\def\expandafter\@authoraddress\expandafter
{\@authoraddress{\nointerlineskip\vskip1pc
                 \footnotesize\it\centering\ignorespaces#1\par}}}

\def\@maketitle{%
\@title
\ifdim\prevdepth=-1000pt \prevdepth0pt\fi
\@authoraddress
}

\def\maketitle{\par
\begingroup
\let\cite\@bylinecite
\global\@topnum\z@ %
\@maketitle
\endgroup
\def\@thanks{}\def\@authoraddress{}\def\@title{}
}

\def\abstract{\par
\bgroup
\ifdim\prevdepth=-1000pt \prevdepth0pt\fi
\hsize\columnwidth
\leftskip=2em \rightskip\leftskip
\dimen0=-\prevdepth \advance\dimen0 by2pc \nointerlineskip
\noindent\vskip1.5\baselineskip\nointerlineskip\noindent\footnotesize\relax}

\newif\if@firststuff

\def\endabstract{\par
\nointerlineskip \vskip0pt
\noindent \par
\egroup
\hrule depth0pt width0pt
\global\everypar{\global\@firststufffalse}\global\@firststufftrue
}

\renewcommand\section{\@startsection {section}{1}{\z@}%
                                   {-3.5ex \@plus -1ex \@minus -.2ex}%
                                   {2.3ex \@plus.2ex}%
                                   {\normalfont\large\bfseries}}
\renewcommand\subsection{\@startsection{subsection}{2}{\z@}%
                                     {-3.25ex\@plus -1ex \@minus -.2ex}%
                                     {1.5ex \@plus .2ex}%
                                     {\normalfont\large\bfseries}}

\def\1ad{\mbox{\normalsize $^1$}}
\def\2ad{\mbox{\normalsize $^2$}}
\def\3ad{\mbox{\normalsize $^3$}}
\def\4ad{\mbox{\normalsize $^4$}}
\def\5ad{\mbox{\normalsize $^5$}}
\def\6ad{\mbox{\normalsize $^6$}}
\def\7ad{\mbox{\normalsize $^7$}}
\def\8ad{\mbox{\normalsize $^8$}}

\makeatother

\usepackage{amsfonts}
\def\be{\begin{eqnarray}}
\def\ee{\end{eqnarray}}
\def\lb{\label}

\begin{document}
\raggedbottom

\title{Black hole production via  quantum tunneling}

\authors{Sergey N.~Solodukhin}
  
\addresses{ Theoretische Physik,
Ludwig-Maximilians Universit\"{a}t, 
Theresienstrasse 37,
D-80333, M\"{u}nchen, Germany}

\maketitle

\begin{abstract}
Particles colliding at impact parameter much larger than the effective 
gravitational radius can not classically form a black hole and just 
scatter off 
the  radial potential barrier
separating the particles. 
We show that the process of the black hole production 
can still go quantum-mechanically
via familiar mechanism of the under-barrier tunneling.
The mechanism is illustrated for  collision of a trans-Planckian particle
and a lighter particle.
Our analysis  reveals instability of trans-Planckian particles against
transition to the phase of black hole.
\end{abstract}

\bigskip

It has been for long time an intriguing problem to study in detail the possibility to create 
 black holes in collision of particles. Many attempts \cite{1}
were done in the past to attack this
problem although the technical difficulties are such that exact solutions
of General Relativity describing this process are still not available.
(Although   a considerable progress in this direction 
has been recently made in \cite{Giddings}).
This problem however became of a practical interest recently due to 
increasing popularity of the models
predicting higher-dimensional Planckian scale at 1 TeV.
If these models are correct the trans-Planckian energy should be available already
in the next generation of colliders or in cosmic rays. 
The black hole production then should be a typical event 
the experimentalists are going to detect \cite{2}.

Although a detailed  theoretical description of the black hole production in trans-Planckian 
collisions is not yet known
some basic principles are available. Mostly, it is based on the picture of the gravitational 
collapse as we know it in General Relativity. In particular, applying  Thorne's hoop
conjecture \cite{TH} to the present situation we arrive at a conclusion
that two particles with energy ${\tt E}_1$ and ${\tt E}_2$ respectively
should form a black hole if they collide 
at impact parameter $b$ less than the effective gravitational
radius of the system, $r_g=2GM$, $M={\tt E}_1+{\tt E}_2$ being the total energy of the particles.
Indeed, they approach each other in the 
spacetime region sufficiently small to turn on the strong gravitational interaction.
On the other hand, for particles approaching each other at shortest distance larger than
$r_g$ the gravitational force is not strong enough and particles just scatter.
The reason for that is simple: the gravitational attraction between the particles 
should compensate the centrifugal repulsion in order to particles could 
come close to each other. The fight of two forces manifests in the radial
potential barrier separating the particles. Low energy particles or particles 
falling towards each other at 
large impact parameter $b$ come close to the bottom of the barrier
and thus get reflected not forming a black hole. 
This situation   is however  the most typical one:
the wave length of particles normally available in experiments is much bigger than
the corresponding gravitational radius. Such particles can not produce black hole
in the classical theory. In this note we  demonstrate that
the process can still go quantum-mechanically via the mechanism of quantum tunneling
through the barrier. Throughout, the standard four-dimensional gravity is understood,
generalizations to higher dimensions being either straightforward or trivial.

To set the stage we need a simple model in which the analysis of the black hole production
could be done. An appropriate 
model was suggested in \cite{SS} and consists in replacing the original
picture of two colliding particles by an approximate one in which a test particle
with energy $\omega={{\tt E}_1 {\tt E}_2\over {\tt E}_1+{\tt E}_2}$ falls onto the gravitating center of
mass $M={\tt E}_1+{\tt E}_2$ from infinity at 
the impact parameter $b$. In the Newtonian mechanics  the second picture is 
completely equivalent to the 
original. It is not the case in General Relativity. A good approximation however
is expected, at least in the case when one of the particles is much heavier than another.
The total gravitational field is modelled by the field of the gravitating center 
and is described by the Schwarzschild metric. It has horizon at $r=r_g$.
We consider it to be  a fictitious horizon 
until it gets crossed by the test particle. 
The test particle then inevitably falls to the center.
This signals for  the formation of  
the black hole (with the actual event horizon being formed) 
in the original picture. The study of black hole production in this model
thus reduces to the
analysis of the gravitational capture of the test particle by the gravitating center.

First, let us analyse the process classically.
The classical radial motion of the test particle is determined 
by the geodesic equation (we put $c=1$)
\be
({dr\over dt})^2=(1-{r_g\over r})^2 b^2\left({1\over b^2}-V(r)\right)~~,~~~ 
V(r)={1\over r^2}(1-{r_g\over r})~~,
\lb{2.1}
\ee
where we used the fact that for a ultra-relativistic
particle, its energy $\omega$ and angle momentum $L$ 
(computed with respect to the gravitating center)
are related as $L/\omega=b$. The potential $V(r)$ in (\ref{2.1}) plays the role of the 
effective
radial potential between the gravitating center and the test particle.
Its form is a result of the interplay of two forces acting in the opposite directions: the
gravitational attraction and the centrifugal repulsion. 
This potential vanishes at horizon, $r=r_g$, and at infinity, $r\rightarrow \infty$,
and develops a maximum in between at $r_m={3\over 2}r_g$.
For a particle coming from infinity at the impact parameter $b$ the crucial relation is 
the relation between $1/b^2$ and the maximal value $V(r_m)={4\over 27 r^2_g}$
of the
effective radial potential.
Particle coming from infinity with
the impact parameter
$b<b_{\tt cr}=3\sqrt{3}/2 r_g$  goes above the maximum of the potential and
gets captured by the gravitating center. 
Having in mind the original
picture of the colliding particles we  say that there forms a black hole with 
size $r_g$.  As a side remark note that in the Newtonian mechanics the radial potential
monotonically grows to infinity when particles approach each other 
so that the gravitational capture is never possible.
The latter thus is an important feature of  General Relativity and is absent in
the Newtonian case.

If the test particle falls below the maximal value of the potential,
this happens for large impact parameter $b>b_{\tt cr}$ , the particle just scatters 
off the center and the gravitational capture does not happen.
In the original picture, this would correspond to particles passing each other 
at far distance without actually forming  a black hole.  
Thus, this model predicts that, classically, two colliding particles 
form a black hole if they pass each other at the shortest distance
$b<b_{\tt cr}=3\sqrt{3}/2 r_g$ (where $r_g$ is the gravitational radius for the
system of these two particles). The particle with $b>b_{\tt cr}$ comes close to the 
bottom of the potential barrier and gets reflected.
Thus, the classical cross-section 
for the black hole formation is given by
\be
\sigma_{\tt cl}=\pi b^2_{\tt cr}={27\over 4} \pi r^2_g=27\pi G^2 M^2~~.
\lb{2.2}
\ee

This picture can  be also  analysed quantum mechanically.
The relevant processes, the black hole scattering and absorption, were well studied
in the past. With new interpretation according to our picture of the black hole production
this study  appears in a new light. 
Decomposing quantum fields in the spherical harmonics, $\Phi_{lm}\sim {u_l(r,\omega )\over r}
e^{-i\omega t}Y_{lm}(\theta, \phi )$, one arrives at the radial wave equation
\be
\left({d^2 \over dr^2_*}+\omega^2-U_l(r)\right) u_l(r,\omega)=0,
\lb{2.3}
\ee
which is a quantum mechanical analog of the classical equation (\ref{2.1}).
We denote ${d\over dr_*}=(1-{r_g\over r}){d\over dr}$ and
\be
U_l(r)=(1-{r_g\over r})\left({l(l+1)\over r^2}+{r_g (1-s^2)\over r^3}\right)~~,
\lb{2.4}
\ee
where $s$ is spin of the test particle (below we put $s=0$),
is the effective potential similar to the potential in eq.(\ref{2.1}).
The equation (\ref{2.3}) describes the scattering by the potential (\ref{2.4}).
The relevant modes are defined to be in- and out-going at infinity $r_*\rightarrow \infty$,
$u_l(r,\omega )\sim A_{\tt out}(\omega )e^{i\omega r_*}+ A_{\tt in}
(\omega )e^{-i\omega r_*}$,
and  only out-going at horizon $u_l(r,\omega )\sim e^{-i\omega r_*}$, 
$r_*\rightarrow -\infty$. Again,  we say that the black hole 
production takes place in the original picture if the wave gets absorbed by the 
fictitious horizon
surrounding the gravitating center of mass $M$.
The probability of a wave to penetrate through the potential barrier is 
$$
\Gamma_{l,\omega}=1-|{A_{\tt out}\over A_{\tt in}}|^2~~.
$$
Note, that it is probability for a spherical wave. Since the actual wave at infinity 
looks more like a plane wave
the latter should be decomposed in the spherical modes
\be
e^{-i\omega z}=\sum_{l=0}^\infty K_{l}(\omega )Y_{l0}(\theta,\phi )~~, ~~~
K_l(\omega )={i^l\over 2\omega} [4\pi (2l+1)]^{1/2}~~.
\lb{2.5}
\ee
The  cross-section to capture the test particle is then defined as follows
\be
\sigma_{\tt quant}=\sum_l |K_l|^2\Gamma_{l,\omega }~~.
\lb{2.6}
\ee
It is also convenient to consider each partial wave characterized by
angular momentum $l$ as falling on the gravitating center at impact
parameter $b$,
$$
b=(l+{1\over 2}){1\over \omega }~~.
$$
The potential (\ref{2.4}) is quite similar to the potential $V(r)$ in
the classical problem (\ref{2.1}). Analyzing the scattering of a partial wave
by this potential it is  important to find relation between
the energy $\omega$ of the in-falling particle and the height of the 
radial potential barrier. For small $b$ (large energy $\omega$ at fixed $l$), one finds that 
the height of the barrier  is the same  as in the classical case and corresponds to the
same impact parameter 
$b=b_{\tt cr}$.
This is what we could have expected: 
in the high energy limit the geometrical optics analysis based on
equation (\ref{2.1}) becomes a good approximation.
In this limit  the horizon absorbs partial waves 
which go above the potential barrier, i.e. 
with the impact parameter $b\leq {3\over 2} r_g$, so that
the cross-section (\ref{2.6})
should approach the classical value (\ref{2.2}).

We are however more interested in the opposite limit
of small $\omega$.
It can be also viewed as limit of small $\omega r_g\ll1$ or, equivalently, of large
impact parameter $b\gg r_g$. 
In this case the test particle falls very close to the bottom of the
potential barrier and, classically, can not penetrate through the
barrier. The quantum mechanical picture however is more intriguing.
The quantum particle can {\it tunnel} through the barrier. Note that in the present case we do
not expect exponentially small factors, usually accompanying the semiclassical analysis of 
the quantum tunneling, to appear. The semiclassical picture fails to describe
the tunneling of small energy quantum particle. 
Rather, we have to rely on exact analysis. Indeed, it is well known from standard courses of 
Quantum Mechanics (see for example \cite{Landau})
that the probability to tunnel through the barrier
grows as  some power of the energy when energy is close to zero.
Specifically, for the problem at hand, the transmission probability 
$\Gamma_{l,\omega }$ grows as  $\omega^{2l+2}$ for small $\omega$ 
and thus dominates for s-wave.
On the other hand, the contribution
of  s-wave in the in-falling plane wave diverges  $|K_l|^2\sim {1\over \omega^2}$,
as is seen from (\ref{2.5}). 
The two tendencies compensate each other in the cross-section (\ref{2.6})
for the s-wave. As a result, in the regime of small $\omega$ the cross-section (\ref{2.6})
approaches a finite number equal to the  horizon 
area \cite{3}
\be
\sigma_{\tt quant}\simeq \pi r^2_g=4\pi G^2M^2~~.
\lb{2.8}
\ee
Since only the wave that approaches  horizon radially 
is absorbed  it is not surprising that the cross-section becomes equal to the
horizon area.
Thus, despite the fact that the process is forbidden classically it still
can go via the mechanism of the quantum tunneling through the potential barrier,
the cross section (\ref{2.8}) being entirely due to
the quantum tunneling effect. 

The discussed mechanism should play an important role in the collision of a trans-Planckian particle
with a lighter  particle. In this case the total energy is
trans-Planckian, ${\tt E}_1+{\tt E}_2\gg{\tt E}_{\tt pl}$, while the size 
of the system given by the
wave length of the lighter particle is much bigger than the effective gravitational
radius of the system, 
i.e. we have that $\omega r_g\simeq {\tt E}_1{\tt E}_2/{\tt E}_{\tt pl}^2\ll 1$.
Classically, the black hole formation is forbidden in this case.
The black hole can nevertheless be produced via the tunneling 
mechanism discussed above.
Our consideration also sheds light on the nature of the trans-Planckian
particles \cite{4}.
Indeed, such a particle is characterized by a wave length $\lambda\sim 1/{\tt E}$
which is much smaller 
than the corresponding gravitational radius $r_g\sim {\tt E}/{\tt E}_{\tt pl}^2$. 
So it is an example of a system which lies
within its gravitational radius. However, we can not call it black hole
since the whole effect is purely kinematic:
one can go to another, boosted, reference frame in which the massive 
particle is at rest
and clearly not a black hole. Our consideration suggests that  
trans-Planckian particles
are unstable against the black hole formation. Indeed, 
the collision with any low energy particle, it meets on its way, sets 
the frame invariant energy 
scale. The low-energy particle then gets absorbed  via the tunneling mechanism just 
described  and the  black hole forms.
Due to this process a trans-Planckian particle which happens to move 
in a medium filled 
with lower-energy particles makes transition to the essence of completely different nature.
Atmosphere of Earth might be such a medium to initiate this 
transition for the
ultra high-energy intruders.

\bigskip

This work is   supported by the  grant DFG-SPP 1096.

\end{document}